\documentclass[showpacs,twocolumn,aps,prl]{revtex4}

\usepackage{amsmath,amssymb}
\usepackage{graphicx}
\usepackage{pspicture}
\usepackage{pst-plot}

\begin{document}

\title{Can Bell's Prescription for Physical Reality Be Considered Complete?}

\author{Joy Christian}

\email{joy.christian@wolfson.ox.ac.uk}

\affiliation{Department of Physics, University of Oxford, Parks Road, Oxford OX1 3PU, United Kingdom}

\begin{abstract}
An experiment is proposed to test Bell's theorem in a purely macroscopic domain. If realized, it would
determine whether Bell inequalities are satisfied for a manifestly local, classical system. It is
stressed why the inequalities should not be presumed to hold for such a macroscopic system without actual
experimental evidence. In particular, by providing a purely classical, topological explanation for the
EPR-Bohm type spin correlations, it is demonstrated why Bell inequalities must be violated
in the manifestly local, macroscopic domain, just as strongly as they are in the microscopic domain.
\end{abstract}


\maketitle

Despite the existence of an explicit counterexample \cite{Disproof}, Bell's theorem
is still widely believed to have proved that no physical theory can be reconciled
with the notion of local reality espoused by Einstein, Podolsky, and Rosen (EPR) \cite{Further}\cite{EPR}.
It therefore seems worthwhile to investigate the very foundations of Bell's theorem
experimentally, in a purely macroscopic domain. If realized, the experiment described
below would test whether or not a manifestly local, macroscopic system can violate Bell
inequalities, as implied by the arguments of Ref.${\,}$\cite{Disproof}. A physical
scenario well suited for this purpose is that of the local model first considered by
Bell himself \cite{Bell-1964}. The details of this model can be found also in some
standard textbooks \cite{Peres-1993}. To our knowledge, no real experiment has ever
been performed to check whether Bell inequalities do indeed hold in the manifestly
local and realistic domain of Bell's model.

The central contention of Ref.${\,}$\cite{Disproof} is that the sinusoidal EPR-Bohm
correlations observed in the laboratory have nothing to do with entanglement or nonlocality
{\it per se}, but stem entirely from the topological properties of the physical space.
This viewpoint can be explained clearly by a closer examination of the model for
spin considered by Bell. In this model the space of complete states of spin consists of
unit vectors ${\boldsymbol{\lambda}}$ in three-dimensional Euclidean space ${{\mathbb E}_3}$.
The local beables ${A_{\bf a}({\boldsymbol\lambda})}$ and ${B_{\bf b}({\boldsymbol\lambda})}$,
existing at freely chosen unit directions ${\bf a}$ and ${\bf b}$, are defined by
\begin{equation}
A_{\bf n}({\boldsymbol\lambda})\,=\,-\,
B_{\bf n}({\boldsymbol\lambda})\,=\,{sign}\,({\boldsymbol\lambda}\cdot{\bf n})\,,
\label{bell-c1}
\end{equation}
provided ${{\boldsymbol\lambda}\cdot{\bf n}\,\not=\,0}$ for ${\bf n=a}$ or ${\bf b}$, and
otherwise equal to the sign of the first nonzero term from ${\{n_x,\,n_y,\,n_z\}}$. This simply
means that ${A_{\bf n}({\boldsymbol\lambda})=+\,1}$ if the two unit vectors ${\bf n}$ and
${\boldsymbol\lambda}$ happen to point through the same hemisphere centered at the origin
of ${\bf n}$, and ${A_{\bf n}({\boldsymbol\lambda})=-\,1}$ otherwise. As a visual aid
to Bell's model \cite{Peres-1993} one can think of a bomb at rest exploding into two
freely moving fragments with angular momenta ${{\boldsymbol\lambda}={\bf J}_1=-{\bf J}_2}$,
with ${{\bf J}_1+{\bf J}_2=0}$. The two functions ${A_{\bf a}({\bf J}_1)}$ and
${B_{\bf b}({\bf J}_2)}$ can then be taken as ${{sign}\,({\boldsymbol\lambda}\cdot{\bf a})}$
and ${{sign}\,(\,-\,{\boldsymbol\lambda}\cdot{\bf b})}$, respectively. If the initial
directions of the two angular momenta are uncontrollable but describable by an isotropic
probability distribution ${\rho({\boldsymbol\lambda})}$ (normalized on the space
${{\mathbb E}_3}$), then, employing the local realistic prescription provided by Bell, namely
\begin{equation}
{\cal E}({\bf a},\,{\bf b})\,=\int_{{\mathbb E}_3}
A_{\bf a}({\boldsymbol\lambda})\,B_{\bf b}({\boldsymbol\lambda})\;\,d\rho({\boldsymbol\lambda}),\label{prob-1}
\end{equation}
the expectation values of the individual variables ${A_{\bf a}({\boldsymbol\lambda})}$ or
${B_{\bf b}({\boldsymbol\lambda})}$ can be easily shown to vanish
identically \cite{Peres-1993}.
Their joint correlation function on the other hand would not vanish in general, and is usually worked out to be
\cite{Peres-1993}
\begin{equation}
{\cal E}({\bf a},\,{\bf b})\,=\,-1+\frac{2}{\pi}\,\cos^{-1}\left({\bf a}\cdot{\bf b}\right).\label{classprob}
\end{equation}
If we now substitute this linear correlation function into the CHSH string of
expectation values for four arbitrarily chosen detector directions ${\bf a}$,
${\bf a'}$, ${\bf b}$, and ${\bf b'}$, giving
\begin{equation}
{\cal E}({\bf a},\,{\bf b})\,+\,{\cal E}({\bf a},\,{\bf b'})\,+\,
{\cal E}({\bf a'},\,{\bf b})\,-\,{\cal E}({\bf a'},\,{\bf b'}),
\label{CHSH-op}
\end{equation}
then it is easy to check that the absolute value of this string never exceeds the
bound of 2, thus saturating but not violating the celebrated Bell-CHSH inequalities \cite{Clauser}.

This often quoted result is usually considered to be well established, but in fact it is simply
incorrect. The trouble is that the local realistic prescription for spin correlations provided
by Bell---namely, Eq.(\ref{prob-1}) above---is incapable of accounting for the elements of
physical reality envisaged by EPR {\it in the topologically correct order}. The situation is analogous
to having taken a photograph apart pixel by pixel, keeping count of each pixel correctly, and then
trying to put it back together. If the geometrical order of the pixels has been neglected in the
process, then there would be little chance of recovering the photograph back. Similarly, what is
missing from the prescription (\ref{prob-1}) is not so much the operational accounting of the
elements of physical reality, but how these elements are coalesced together topologically. As
we shall see, the correct result for the spin correlations, derived using both operationally and
topologically complete prescription, works out to be
\begin{equation}
{\cal E}({\bf a},\,{\bf b})\,=\,-\,{\bf a}\cdot{\bf b}\,,\label{cliffordprob-1}
\end{equation}
which extends the bound on the Bell-CHSH inequality from 2 to ${2\sqrt{2}}$.
To fully understand this classical result let us take a closer look at the
derivation of Eq.(\ref{classprob}). 

Since the initial distribution of the angular momenta is supposed to have been isotropic, in
Bell's model the space of all possible directions of ${{\bf J}_1}$---that is, both the configuration space as well as
the phase space of ${{\bf J}_1}$---is traditionally \cite{Peres-1993} taken to be a unit 2-sphere, defined by
\begin{equation}
n^2_x+n^2_y+n^2_z=1.\label{2-seven}
\end{equation}
Next, since each point ${\boldsymbol\lambda}$ on this surface represents an EPR element of physical reality,
the integration in Eq.(\ref{prob-1}) is carried out over this surface, yielding the result (\ref{classprob}).
Now the group of linear transformations that leave such a quadratic form invariant is the orthogonal
group O(3), which includes both rotational and reflective symmetries of the 2-sphere. On the other hand,
we know that angular momentum is not an ordinary polar vector, but a {\it pseudo} vector that changes sign
upon reflection. One only needs to compare a spinning object with its image in a mirror to confirm this
fact. This familiar fact is sufficient, however, to divulge the first sign of trouble with Bell's chosen
set of observables---namely, ${{sign}\,({\boldsymbol\lambda}\cdot{\bf n})}$. Clearly, since ${\boldsymbol\lambda}$
is supposed to be the spin angular momentum whereas ${\bf n}$ is simply an ordinary polar vector,
the dot product in the observable ${{sign}\,({\boldsymbol\lambda}\cdot{\bf n})}$ cannot be a {\it true}
scalar, but a {\it pseudo} scalar---one that changes sign in the mirror.

There is of course an easy way out of this problem. All one has to do is to restrict the symmetry group of
the 2-sphere to the subgroup SO(3)---i.e., to the group of non-reflective symmetries. This seems straightforward
enough, but one must bear in mind that, although their Lie algebras are identical, globally the groups O(3) and SO(3)
are profoundly different from each other. Globally the group SO(3) is a highly non-trivial subgroup of O(3). Indeed,
topologically the space SO(3) is homeomorphic to the real projective space ${{\mathbb R}{\mathbb P}^3}$, which is
a connected, but not simply-connected manifold \cite{Frankel}\cite{Penrose}. That is to say, there are loops in SO(3)
that cannot be contracted to a point. In physical applications this fact is well known to give rise to unavoidable
singularities, discontinuities, and wildly spinning trajectories \cite{Heard}. In addition to this fatal defect,
the group SO(3) also harbors a related conceptual defect, which is of profound significance for our concerns. The
trouble is that SO(3) does not always respect the true rotational symmetries of the physical space.

To appreciate this well known fact, consider a rock in an otherwise empty universe. If such a rock is allowed
to rotate by ${2\pi}$ radians about some axis, then it will return back to its original state. This, however,
will not happen if there is at least one other object present in the universe. The rock will then have to 
rotate by another ${2\pi}$ radians (i.e., a total of ${4\pi}$ radians) to return back to its original
state, relative to that other object. This well known fact is often demonstrated by a ``belt trick''
(cf. \cite{Penrose}, p 205),
which shows that what is an identity transformation for an isolated object is {\it not} an identity
transformation for an object that is rotating in the presence of other objects.
Thus, what appears to be an identity transformation in the latter case on purely operational basis,
is simply an illusion. This peculiar property of the ordinary objects is not
respected by the structure of SO(3). That is to say, SO(3) is capable of providing only tensor
representations of the rotation group, and not its spinor representations. This is fine as long as one
is concerned with rotations of only isolated objects, but it is anything but fine in our case, since we
are concerned with correlations between two macroscopic bomb fragments {\it rotating in tandem}.

Fortunately \cite{Frankel}\cite{Heard}, all of the above difficulties can be resolved by representing rotations
in physical space by elements of the universal covering group of SO(3), namely the group SU(2) of unit
quaternions (or spinors \cite{Penrose}, or rotors \cite{Hestenes}). This group can be constructed by taking
{\it two} copies of SO(3), and gluing their boundaries together point by point, so that each ${-\pi}$
rotation-point on the boundary of one copy is identified with the respective ${+\pi}$ rotation-point on
the boundary of the second copy. The resulting space is a topological 3-sphere defined by
\begin{equation}
n^2_o+n^2_x+n^2_y+n^2_z=1,\label{3-seven}
\end{equation}
where the quadruple ${(\,n_o,\,n_x,\,n_y,\,n_z)}$ defines a non-pure unit quaternion \cite{Frankel}.
The 3-sphere is well known of course to have exceptionally special properties \cite{sphere}. It is the only
three-dimensional manifold without boundary that is not only compact and connected, but also simply-connected. And
it is the only simply-connected, parallelizable sphere that is homeomorphic to a Lie group, namely SU(2) (it is also
worth noting the obvious that the usefulness of this group is not exclusive to quantum mechanics \cite{Heard}).

\begin{figure}
\hrule
\scalebox{0.60}{
\begin{pspicture}(0.3,-3.7)(4.2,2.8)

\pscircle[linewidth=0.3mm,linestyle=dashed](-1.8,-0.45){2.6}

\psellipse[linewidth=0.3mm](-0.8,-0.45)(0.7,1.4)

\psellipse[linewidth=0.3mm,border=3pt](-2.4,-0.45)(1.4,0.4)

\pscurve[linewidth=0.3mm,border=3pt](-1.485,-0.35)(-1.48,-0.25)(-1.45,0.0)

\pscircle[linewidth=0.3mm](7.0,-0.45){1.7}

\psellipse[linewidth=0.2mm,linestyle=dashed](7.0,-0.45)(1.68,0.4)

\put(-4.4,1.27){{\Large ${S^3}$}}

\put(-2.0,1.2){{\Large ${h^{-1}(q)}$}}

\put(-3.7,-1.4){{\Large ${h^{-1}(p)}$}}

\put(7.43,0.67){{\Large ${q}$}}

\psdot*(7.2,0.79)

\put(6.3,0.43){{\Large ${p}$}}

\psdot*(6.1,0.43)

\put(8.5,-1.8){{\Large ${S^2}$}}

\put(5.9,-2.7){\Large base space}

\put(1.7,0.7){\Large ${h:S^3\rightarrow S^2}$}

\pscurve[linewidth=0.3mm,arrowinset=0.3,arrowsize=3pt 3,arrowlength=2]{->}(1.2,0.25)(2.47,0.45)(3.74,0.45)(4.9,0.25)

\put(1.8,-0.45){\large Hopf fibration}

\pscurve[linewidth=0.3mm,arrowinset=0.3,arrowsize=3pt 3,arrowlength=2]{->}(4.9,-0.95)(3.74,-1.15)(2.47,-1.15)(1.2,-0.95)

\put(1.55,-1.8){\Large ${h^{-1}:S^2\rightarrow S^3}$}

\end{pspicture}}
\hrule
\caption{The tangled web of linked Hopf circles depicting the topological intricacies of the EPR elements of
physical reality.}
\label{fig}
\smallskip
\hrule
\end{figure}
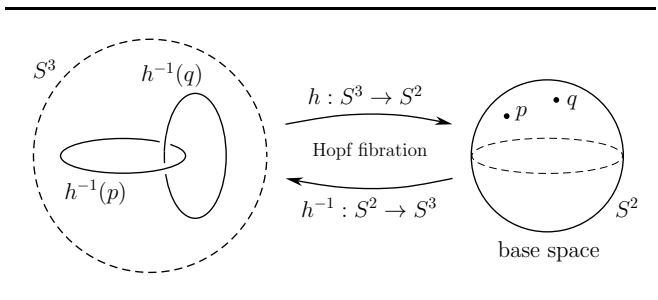

Despite its being contained in ${{\mathbb R}^4}$, it is in fact possible to ``see'' inside this sphere by means
of a Hopf fibration \cite{Ryder}. This provides us an opportunity to appreciate the true topological structure
of the elements of reality for our bomb fragments. As illustrated in Fig.${\,}$\ref{fig}, the 2-sphere we
started out with, namely the one defined by Eq.(\ref{2-seven}), turns out to be only the base manifold of this
profound structure. The points of this base manifold, namely ${S^2}$, now correspond to elements of the Lie
algebra su(2), and are in fact {\it pure} quaternionic numbers \cite{Frankel}\cite{Penrose}. The product of
two such numbers on ${S^2}$ are then {\it general}${\,}$ quaternionic numbers, defined by (\ref{3-seven}), and
belong to the group SU(2) itself. That is to say, they are points on the bundle space ${S^3}$, which is
{\it completely}${\,}$ made up of the preimages of the points on the base ${S^2\,}$\cite{Ryder}.
${\!}$These preimages are 1-spheres, ${S^1}$, called Hopf circles, or Clifford parallels
(\cite{Penrose}, p 335). Since these 1-spheres are the fibers of the bundle, they do not share a single
point in common. And yet each circle threads through every other circle in the bundle, making them all linked
together in a highly intricate fashion. In particular, although locally the bundle ${S^3}$ is a product
space ${S^2\times S^1}$, {\it globally it has no cross-section at all}.

It should be fairly clear by now that topologically the EPR elements of reality have far
deeper structure than has been hitherto appreciated. Clearly, no prescription that ignores
this structure can be expected to provide the correct correlation function for our bomb fragments.
In particular, no Bell-type scalar functions of the form
\begin{align}
&A_{\bf n}(\lambda): V\!\times\Lambda\longrightarrow S^0\equiv\{-1,\,+1\}, \\
{\rm with}\;\;
A_{\bf a}&(\lambda)B_{\bf b}(\lambda)\,\in\, S^0\times S^0 \,=\, S^0\equiv\{-1,\,+1\}
\end{align}
(where ${V}$ is a vector space and ${\Lambda}$ is a space of ``complete'' states),
can account for the topological intricacies of the elements of physical reality,
{\it even for our purely classical rotors}. Surely, no elements of a 0-sphere can
imitate the topological profundities of the elements of a 3-sphere.

On the other hand, from the above picture, and from the fact that the groups O(3), SO(3), and SU(2)
all share the same Lie algebra structure, it is clear that simply promoting the variables 
${A_{\bf n}({\boldsymbol\lambda})}$ to be the elements of Lie algebra su(2) cannot be sufficient
to capture the global, topological features of the group SU(2). Capturing these features is mandatory,
however, if we are to represent the EPR elements of physical reality faithfully within the choice
of our dynamical variables. Hence, recalling the definition of a group, what we must
ensure is not only that the local variables---which we write as ${A_{\bf n}({\boldsymbol\mu})}$---are
functions of the elements of the Lie algebra su(2), but that their group products,
${A_{\bf a}({\boldsymbol\mu})\,B_{\bf b}({\boldsymbol\mu})}$, appearing in the integrand of
Eq.(\ref{prob-1}), are themselves genuine elements of the group SU(2); for what is captured by the Lie
algebra su(2) is only the tangent structure of the group SU(2).
At the same time, we must also ensure of course that the revised variables ${A_{\bf n}({\boldsymbol\mu})}$ 
remain {\it operationally identical} to the original variables ${A_{\bf n}({\boldsymbol\lambda})}$.
We may then have a chance of evaluating their correlations correctly, by employing an
appropriately generalized expectation functional \cite{Segal}. In sum, the necessary and sufficient
conditions on the local beables for obtaining the correct correlation function are:
\begin{align}
{\rm su(2)}\,\ni\;&A_{\bf n}({\boldsymbol\mu})\,=\,\pm1 \;\, {\rm about} \;\, {\bf n}, \\
{\rm and}\;\;\,{\rm SU(2)}\,\ni\,A_{\bf a}&({\boldsymbol\mu})\,B_{\bf b}({\boldsymbol\mu})\,=\,{\rm a\;unit\;quaternion}.
\end{align}

A local realistic model for the EPR-Bohm correlations satisfying precisely these conditions has been proposed
in Ref.\cite{Disproof}. The complete state specifying all of the elements of reality in this model
is taken to be the unit trivector
\begin{equation}
{\boldsymbol\mu}\,=\,{\bf u}\,\wedge\,{\bf v}\,\wedge\,{\bf w}\,=\,\pm\,I\,\equiv\,
\pm\;{{\bf e}_x}\,\wedge\,{{\bf e}_y}\,\wedge\,{{\bf e}_z}\,,\label{tri-mu}
\end{equation}
where ${\bf u}$, ${\bf v}$, and ${\bf w}$ are vectors of arbitrary length, and ${I}$ is the fundamental
volume form on the physical space. The specification of the complete state ${\boldsymbol\mu}$ predetermines
the entire geometry of the three-dimensional Euclidean space ${{\mathbb E}_3}$ (encapsulated in the Clifford
algebra ${Cl_{3,0}}$). It determines all scalars by their duality relations with ${\boldsymbol\mu}$, all
vectors ${\bf x}$ by definition ${{\boldsymbol\mu}\wedge{\bf x}=0}$, all bivectors by the duality relation
${{\boldsymbol\mu}\cdot{\bf x}={\boldsymbol\mu}\,{\bf x}}$, and all quaternions by the Clifford product
${\;{\bf x}\,{\bf y}={\bf x}\cdot{\bf y}+{\boldsymbol\mu}\cdot({\bf x}\times{\bf y})}$. The locally specified
beables of the model are then taken to be the unit bivectors ${{\boldsymbol\mu}\cdot{\bf n}}$,
which are elements of the Lie algebra su(2) of the group SU(2), with the following properties:
\begin{align}
&\;{\boldsymbol\mu}\cdot{\bf n}=\pm\,1\;\,\text{about the dual vector}\;\,{\bf n}, \\
\text{and}\;\,&(\,{\boldsymbol\mu}\cdot{\bf a})\,(\,{\boldsymbol\mu}\cdot{\bf b})
=\,-\,{\bf a}\cdot{\bf b}\,-\,{\boldsymbol\mu}\cdot({\bf a}\times{\bf b}).
\end{align}
Note that the Clifford product ${({\boldsymbol\mu}\cdot{\bf a})({\boldsymbol\mu}\cdot{\bf b})}$ within 
${Cl_{3,0}}$ is also a group product within SU(2), yielding a non-pure unit quaternion \cite{Frankel}\cite{Ryder}.
This can be verified by comparing the decomposition of the above product with Eq.(\ref{3-seven}).
A generalized expectation functional analogous to (\ref{prob-1}) then gives the correct correlation
function for our rotors:
\begin{equation}
{\cal E}({\bf a},\,{\bf b})=\int_{{\cal V}_3}
(\,{\boldsymbol\mu}\cdot{\bf a}\,)
(\,{\boldsymbol\mu}\cdot{\bf b}\,)\;\,d{\boldsymbol\rho}({\boldsymbol\mu})
=\,-\,{\bf a}\cdot{\bf b},\label{derive}
\end{equation}
where the integral is a {\it directed} integral, ${{\cal V}_3}$ is a manifold whose ``points'' are {\it vectors}
in ${{\mathbb E}_3}$, and the distribution ${{\boldsymbol\rho}(\boldsymbol\mu)}$ is assumed to be
normalized on this {\it vector} manifold. Clearly, unlike Eq.(\ref{prob-1}), the above
prescription is not only {\it operationally complete}, but also {\it topologically complete}. 
As shown in Ref.${\,}$\cite{Further}, if we now substitute this
correlation function into the CHSH string of expectation values, then the bound on its absolute
value is extended to ${2\sqrt{2}}$.

It is worth noting here that the model described above is not only {\it manifestly realistic}, but also
{\it intrinsically local}. There are several independent ways to verify the latter fact \cite{Further}.
To begin with, it
is evident from their bivectorial constitution that the two remote beables ${{\boldsymbol\mu}\cdot{\bf a}}$
and ${{\boldsymbol\mu}\cdot{\bf b}}$ have nothing to do with each other. In fact, since they are two genuine
elements of the Lie algebra su(2), they are necessarily two independent points on the corresponding
2-sphere. Moreover, as rigorously proved in Ref.${\,}$\cite{Further}, the above model
satisfies not only the condition of parameter independence, but also that of outcome independence.

The central message of Refs.${\,}$\cite{Disproof} and \cite{Further} and the above discussion
is that EPR-Bohm correlations have nothing to do with entanglement or non-locality {\it per se},
but are a vestige of geometry and topology of the physical space. This recognition almost immediately
leads to prediction (\ref{derive}), which differs from the prediction (\ref{classprob}) derived
on the basis of Bell's prescription (\ref{prob-1}). These two predictions are clearly distinguishable.
The experiment described below to distinguish them is essentially a realization of Bell's own local model
discussed above \cite{Peres-1993}.
It can be performed either in the outer space or in a terrestrial laboratory. In the latter
case, however, the effects of gravity and air resistance would complicate matters. For simplicity we shall
assume that experimental parameters can be chosen sufficiently carefully to compensate such effects.

With this assumption, consider a ``bomb'' made out of a hollow toy ball of diameter, say, three centimeters.
The thin hemispherical shells of uniform density that make up the ball are snapped together at their rims
in such a manner that a slight increase in temperature would pop the ball open into its two constituents
with considerable force \cite{Peres-1993}. A small lump of density much grater than the density of the ball
is attached on the inner surface of each shell at a random location, so that, when the ball pops open,
not only would the two shells propagate with equal and opposite linear momenta orthogonal to their common
plane, but would also rotate with equal and opposite spin momenta about a random axis in space. The volume of
the attached lumps can be as small as a cubic millimeter, whereas their mass can be comparable to the mass of
the ball. This will facilitate some ${10^6}$ possible spin directions for the two shells, whose outer surfaces
can be decorated with colors to make their rotations easily detectable.

Now consider a large ensemble of such balls, identical in every respect except for the relative locations of the
two lumps (affixed randomly on the inner surface of each shell). The balls are then placed over a heater---one at a
time---at the center of an EPR-Bohm type setup \cite{Clauser}, with the common plane of their shells held perpendicular
to the horizontal direction of the setup. Although initially at rest, a slight increase in temperature of each
ball will eventually eject its two shells towards the observation stations, situated at a chosen distance in the
mutually opposite directions. Instead of selecting the directions ${\bf a}$ and ${\bf b}$ for observing spin
components, however, one or more contact-less rotational motion sensors---capable of determining the precise
direction of rotation---are placed near each of the two stations, interfaced with a computer. These
sensors will determine the exact direction of the angular momentum ${{\boldsymbol\lambda}_j}$
(or ${-{\boldsymbol\lambda}_j}$) for each shell, without disturbing them otherwise, at a designated distance from
the center. The interfaced computers can then record this data, in the form of a 3D map of all such directions.

Once the actual directions of the angular momenta for a large ensemble of shells on both sides are fully recorded,
the two computers are instructed to randomly choose the reference directions, ${\bf a}$ for one station and ${\bf b}$ for
the other station---from within their already existing 3D maps of data---and then calculate the corresponding dynamical
variables ${{sign}\,({\boldsymbol\lambda}_j\cdot{\bf a})}$ and ${{sign}\,(\,-\,{\boldsymbol\lambda}_j\cdot{\bf b})}$.
This ``delayed choice'' of ${\bf a}$ and ${\bf b}$ will guarantee that the conditions of parameter independence and
outcome independence are strictly respected within the experiment \cite{Further}. It will ensure, for example, that the
local outcome ${{sign}\,({\boldsymbol\lambda}_j\cdot{\bf a})}$ remains independent not only of the remote parameter
${\bf b}$, but also of the remote outcome ${{sign}\,(-{\boldsymbol\lambda}_j\cdot{\bf b})}$. If in any doubt, the two
computers can be located at a sufficiently large distance from each other to ensure local causality while selecting
${\bf a}$ and ${\bf b}$. The correlation function for the bomb fragments can then be calculated using the formula
\begin{equation}
{\cal E}({\bf a},\,{\bf b})\,=\,\frac{1}{N}\sum_{j\,=\,1}^{N}\,
\{{sign}\,({\boldsymbol\lambda}_j\cdot{\bf a})\}\,
\{{sign}\,(-{\boldsymbol\lambda}_j\cdot{\bf b})\},\label{correlations}
\end{equation}
where ${N}$ is the number of trials. This result, which would give purely local correlations,
should then be compared (in ${N\rightarrow\infty}$ limit) with the predictions (\ref{classprob}) and (\ref{derive}).

It is worth recalling here that the variables ${{sign}\,({\boldsymbol\lambda}\cdot{\bf n})}$ and
${{\boldsymbol\mu}\cdot{\bf n}}$ used in the respective derivations of equations (\ref{classprob}) and
(\ref{derive}) are {\it operationally identical} to each other:
\begin{equation}
{sign}\,({\boldsymbol\lambda}\cdot{\bf n})\,\cong\,\pm\,1\;\,{\rm about}\;\,{\bf n}\,\cong\,{\boldsymbol\mu}\cdot{\bf n}.
\end{equation}
This can be easily verified by noting that the variables ${{sign}\,({\boldsymbol\lambda}\cdot{\bf n})}$ are simply
the normalized components of the angular momenta ${{\bf J}}$ along the directions ${\bf n}$, and so are the
variables ${{\boldsymbol\mu}\cdot{\bf n}}$ (albeit in the bivector basis \cite{Hestenes}). In other words,
although mathematically ${{sign}\,({\boldsymbol\lambda}\cdot{\bf n})}$ and ${{\boldsymbol\mu}\cdot{\bf n}}$
are elements of two different grades in the algebra ${Cl_{3,0}}$ (one is a scalar and the other a
bivector), physically they represent one and the same rotor quantity \cite{Further}.

Undoubtedly, there would be many different sources of systematic errors in an experiment such as this.
If it is performed carefully enough, however, then---in the light of the discussion above---we believe the
experiment will vindicate prediction (\ref{derive}) and refute prediction (\ref{classprob}).

\smallskip

I am grateful to Simon Saunders for raising the issue of macroscopic violation of Bell inequalities,
and to Abner Shimony for correspondence concerning Refs.${\,}$\cite{Disproof} and \cite{Further}.

\end{document}